\documentclass[journal]{IEEEtran}

\usepackage{cite}
\usepackage{multicol,lipsum}
\usepackage{amsmath,amssymb,amsfonts}
\usepackage{algorithmic}
\usepackage{graphicx}
\usepackage{textcomp}
\usepackage{xcolor}
\usepackage{booktabs}
  \usepackage{hyperref}
\usepackage{enumerate}
\usepackage{algorithm}
\usepackage{multirow}

\begin{document}
%

 \title{ A Novel  Multi-Task Learning Empowered Codebook Design  for  Downlink SCMA  Networks}
%

%

\author{Qu Luo, ~\IEEEmembership{Student Member,~IEEE,}
        Zilong Liu, ~\IEEEmembership{Senior Member,~IEEE,}
        Gaojie Chen, ~\IEEEmembership{Senior Member,~IEEE,}\\
          Yi Ma, ~\IEEEmembership{Senior Member,~IEEE,}
        Pei Xiao,~\IEEEmembership{Senior Member,~IEEE.}
\thanks{ Qu  Luo, Gaojie Chen, Yi Ma   and  Pei  Xiao  are  with  5G \& 6G  Innovation Centre, University of Surrey, UK, email:\{q.u.luo,  gaojie.chen, y.ma, p.xiao\}@surrey.ac.uk. Zilong   Liu   is   with   the   School   of   Computer   Science   and   Electronics   Engineering,   University   of   Essex,   UK. email:   zilong.liu@essex.ac.uk. This work  was
supported in part by the UK Engineering and Physical Sciences Research Council under Grant EP/P03456X/1 and the Key Project of Science and Technology of Hainan (N0. ZDKJ2019003).}
 }
%
%


\maketitle
\begin{abstract}
Sparse code multiple access (SCMA) is a promising code-domain non-orthogonal multiple access (NOMA) scheme for the enabling of massive machine-type communication. In SCMA,  the design of good sparse codebooks and efficient multiuser decoding have attracted tremendous research attention in the past few years. This paper aims to leverage deep learning to jointly design the downlink SCMA encoder and decoder with the aid of autoencoder. We introduce a novel end-to-end learning based SCMA (E2E-SCMA) design framework, under which improved sparse codebooks and low-complexity decoder are obtained. Compared to conventional SCMA schemes, our numerical results  show that the proposed E2E-SCMA leads to significant improvements  in terms of error rate and computational complexity.
\end{abstract}


\IEEEpeerreviewmaketitle

\section{Introduction}
 \IEEEPARstart{T }{he}  wireless networks are rapidly evolving towards a paradigm shift from connecting people to networking everything. A pressing challenge of future wireless network design is how to develop a highly efficient  multiple access scheme to meet various  stringent requirements such as lower access latency, and higher spectral efficiency.  A disruptive technique for addressing such a challenge  is called non-orthogonal multiple access (NOMA).  In a NOMA system, multiple users are able to communicate simultaneously to achieve overloading factor larger than 1.  Existing NOMA techniques can be largely categorized into two classes: power-domain NOMA  and code-domain NOMA (CD-NOMA)  \cite{liu2021sparse,luo2021error}. In this paper, we focus on an emerging CD-NOMA scheme called sparse code multiple access (SCMA) in which multiple users are separated by adopting different sparse codebooks \cite{nikopour2013sparse,NovelHoshyar}. Over the past decade,  SCMA has attracted tremendous research attention from both academia and industry \cite{chen2020design,yu2015optimized,lai2021analyzing,yu2021sparse2}.

In SCMA, two fundamental research problems are the design of good sparse codebooks and  efficient multi-user decoding \cite{yang2016low,mheich2018design,chen2020design,yu2015optimized }.
Existing known SCMA codebook constructions mostly follow a multi-stage sub-optimal design for rapid generation  \cite{mheich2018design,yu2015optimized,chen2020design}, albeit it is unclear how far the obtained SCMA codebooks are from the optimal ones.
  By taking advantage of the codebook sparsity, low-complexity MPA has been developed for SCMA decoding. For a downlink SCMA system where multiple user devices (e.g., sensors, tablets, machines) are constrained by their limited computation capability and battery life, however, the current MPA may not be affordable, especially when a large number of MPA iterations is needed \cite{yang2016low,lin2020novel}.

In recent years, deep learning (DL)  has  been extensively studied in    wireless networks, thanks to   its capability in solving very complicated optimization problem  \cite{o2017introduction}.
 A comprehensive introduction  on  autoencoder  for end-to-end communication system was contributed  by O'shea and Hoydis in \cite{o2017introduction}.
Following   \cite{o2017introduction}, a denoising autoencoder (DAE) for SCMA  was  reported in \cite{kim2018deep}. The core idea of \cite{kim2018deep} is to  model  the entire SCMA system as a DAE by implementing both the  encoder and decoder with fully connected neural networks (NNs). Subsequently, a similar structure was studied in \cite{lin2020novel,Minsig} by jointly  considering the sparse and dense mapping of CD-NOMA. It is noted that  \cite{kim2018deep,lin2020novel,Minsig} considered  the decoder as a single learning task implemented with fully connected layers.   
However, the bit error performances of these systems may not beat an SCMA system with the aforementioned sparse codebooks that are obtained from a multi-stage sub-optimal design.   Very  recently,  a deep neural network (DNN) with multi-task structure was proposed in  \cite{9310294}  for SCMA detection.
However,  \cite{9310294} has not touched the sparse codebook design with the aid of DNN, and hence a good error rate performance may not be guaranteed.

 In this letter, we introduce a novel multi-task  learning empowered end-to-end SCMA (E2E-SCMA) design framework.  The main novelty of this work stems from  the proposed architecture of E2E-SCMA and the unique   training scheme. Building upon a new  SCMA mapping design with linear encoding, we first propose an  efficient SCMA encoder, which can reduce the   depth of the network and thereby helping  prevent the gradient from vanishing.   Unlike existing works \cite{kim2018deep,lin2020novel,Minsig}, where the decoding is conducted   by viewing $J$ users as a single learning task,  we view each user as a single learning task and then design the decoder in a task-specific fashion.  The advantages of using  the multi-task learning structure are twofold: 1) it can  improve learning efficiency and reduce over-fitting \cite{zhang2017survey}; 2)  it can avoid the curse of dimensionality while  using one-hot encoding.  Specifically, for an multi-task learning structure of $J$ tasks, if each task has a $M$-dimensional input vector, the corresponding input dimension of single  task learning structure will increase to  $M^J$.  Finally, we  propose to train the E2E-SCMA in a  range of signal-to-noise ratios (SNRs) instead of over a fixed SNR. Consequently, this enables the proposed    E2E-SCMA  to work over a wide range of SNR values with a low error rate performance.
 The remainder of the letter is organized as follows. Section II briefly describes the system model of SCMA. We present the proposed E2E-SCMA framework in Section III. The numerical results and conclusion are presented in Sections  IV and V, respectively.

 \vspace{-0.2cm}
\section{System Model}

In this paper, we consider a downlink SCMA system with $J$ users communicating over the  $K$ orthogonal resources, where $J > K$. Let us define  the overloading factor   as $ \lambda = \frac{J}{K}> 1 $. At  the transmitter side,  the SCMA encoder maps $\log_2\left(M\right)$   binary bits toa length-$K$ codeword drawn from codebook $\mathcal {X}_{j}  \in \mathbb {C}^{K}$ with size $M$. The mapping process   is defined as
 $f_j:\mathbb{B}^{\log_2M}  \rightarrow {\mathcal X}_{j}   \in \mathbb {C}^{K}$,  where $\mathcal {X}_{j}=\{\mathbf{x}_{j,1}, \mathbf{x}_{j,2},\ldots,\mathbf{x}_{j,m}\} $   is the codebook set for the $j$th user with cardinality of $M$.
All the $K$-dimensional complex codewords of each  SCMA codebook are sparse vectors with $N$ non-zero elements\footnote{For user $j$, the $N$ non-zero element positions remain unchanged from one codeword to another.} and $N < K$.
Let $\mathbf {c}_{j}$ be a length-$N$ vector drawn from  $  {\mathcal C}_{j}\subset \mathbb {C}^{N }$, where $   {\mathcal C}_{j}$ is obtained  by removing all the zero elements  in  $ { \mathcal X}_{j}$.  We further define the mapping from $\mathbb{B}^{\log_2M}$ to  $ {\mathcal C}_{j}$ as
\begin{equation} \label{SCMAmapping}
\small
g_{j}:\mathbb {B}^{\log _{2}M\times 1}\mapsto   \boldsymbol {{\mathcal C}}_{j}, \quad {~\text {i.e., }}\mathbf {c}_{j}=g_{j}(\mathbf {b}_{j}),
\end{equation}
where $\mathbf {b}_{j}=[b_{j,1},b_{j,2},\ldots,b_{j,\log _{2} M}]^{T}\in \{1,-1\}^{\log _{2} M}$ stands for $j$th user's  instantaneous input binary message vector. By collecting all the  $\mathbf {b}_{j}$ according to their corresponding integer values in ascending order, we form a $\log_2(M) \times M$ binary matrix $\mathbf{B}$.  For example, when $M=4$,  we have
\begin{equation}
\begin{aligned} \mathbf {B}=\left [{ \begin{matrix} -1&+1&-1&+1\\ -1&-1&+1&+1 \end{matrix} }\right].
\end{aligned}
\end{equation}

Thus, the corresponding  SCMA mapping  $f_j$ can be expressed as
\begin{equation}
\label{scmaMapping}
\small
f_{j}:\equiv \mathbf {V}_{j}g_{j}, \quad {~\text {i.e., }}\mathbf {x}_{j}=\mathbf {V}_{j}g_{j}(\mathbf {b}_{j}),
\end{equation}
where $\mathbf {V}_{j} \in \mathbb {B}^{K \times N} $ is an mapping   matrix that maps the $N$-dimensional vector  to a $K$-dimensional sparse SCMA codeword. The sparse structure of the $J$ SCMA codebooks    can be represented by the indicator (sparse) matrix  $\mathbf {F} = \left [ \mathbf {f}_1, \ldots, \mathbf {f}_J \right] \subset \mathbb {B}^{K\times J}$ where  $\mathbf {f}_j = \text {diag}(\mathbf {V}_j\mathbf {V}_j^T)$.

 For a fixed $\mathbf V_j$, the task of  SCMA codebook design is to find   the $J$ mapping functions  $g_{j}, j= 1,2,\ldots,  J$,  according to  certain criteria,  such as minimum Euclidean distance (MED). Specifically, by viewing the  mapping function  $g_{j}$  as a ${N\times \text{log}_2 M}$ complex codebook generator matrix times  the $j$th user's bit vector $\mathbf{b}_j$, we have
\begin{equation} \label{design}
\small
\mathbf {c}_{j} =  {{\mathbf{G}}_{j}}{{\mathbf{b}}_{j}},
 \end{equation}
 where ${{\mathbf{G}}_{j}}\in {{\mathbb C}^{N\times \text{log}_2M}}$ is the codebook generator matrix of the  $j$th user. Therefore, the codebook for user $j$ is  $\mathcal {X}_{j}= \mathbf {V}_{j} {{\mathbf{G}}_{j}}{{\mathbf{B}} }$.

 The received signal of user $j$ in downlink channel   after the multiplexing can be expressed as
\begin{equation}
\small
\begin{aligned}
\mathbf{r}_j =  \operatorname{diag}\left(\mathbf{h}_j\right)  \sum_{u=1}^{J} \mathbf {V}_{u} \mathbf{G}_{u} \mathbf{b}_{u} +\mathbf{n}_j,
\end{aligned}
\end{equation}
where ${{\mathbf{h}}_u}={{\left[ {{h}_{j,1}},{{h}_{j,2}},\ldots, {{h}_{j,K}} \right]}^{T}} \in {{\mathbb{C}}^{K\times 1}}$ is the channel coefficient vector between the base station and  the  $j$th user, diag$ (\cdot ) $ denotes the diagonalization of a matrix and  ${{\mathbf{n}_j} = \left [{ {n_{j,1},{n_{j,2}},\ldots ,{n_{j,K}}} }\right ] }^{T}$ is the complex Gaussian vector with the variance with zero mean and variance $N_{0}$, i.e., ${{n}_{j,k}} \sim \mathcal{CN}\left ({0, {N_{0} ^{}} }\right )$.

In the next section, we will design the near optimal  generating    matrices  $\mathbf{G}_j, j= 1,2,\ldots,  J$ to improve the error rate performance with the proposed novel  autoencoder.

 \vspace{-0.3cm}
\section{Proposed novel autoencoder}
In this section,  a novel autoencoder is presented for   downlink SCMA systems.    We first present the SCMA mapping, i.e., the  signal model in  (\ref{design}),  inspired encoder  designed. Then, the multi-user detection with deep multi-task learning is elaborated. In addition, training procedure and complexity analysis will be discussed.
 \vspace{-0.3cm}
\subsection{Autoencoder}

Autoencoder is a special class  of neural networks which is trained to produce an output data that  matches with their  input data.
It is  composed of a basic  DNN   unit formed of multiple repetitive hidden layers. Each hidden layer is an affine mapping followed by a nonlinearlity operator. The output of the  $l$th hidden layer is given by
    \begin{equation} \label{outl}
\mathbf {x}_{l} =  \sigma_{l} \left( \mathbf {W}_{l} \mathbf {x}_{l-1}+ \mathbf {z}_{l} \right),
    \end{equation}
where   $\boldsymbol {W}_{l}$, $\mathbf{z}_{l}$, and $\sigma_l$ denote  the weight matrix,  bias vector and   the activation function for the $l$th layer, respectively.
The encoder  first  transforms the input vector $\mathbf x$ into hidden representation $\mathbf y$ through a  deterministic mapping $ e_{\boldsymbol {\theta}}$, i.e., $\mathbf y= e_{\boldsymbol{\theta}} \left( \mathbf {x}\right)$, where  $\boldsymbol{\theta}$ denotes the parameter set with all the weight matrices and  bias vectors.
The resulting representation $\mathbf y$ is then mapped back to reconstruct the input vector, i.e., $\mathbf{\hat{x}}= d_{\boldsymbol{\theta}'} \left( \mathbf {y }\right)$. The mapping $d_{{\boldsymbol \theta}}$ is called decoder and $\mathbf {{\boldsymbol \theta'}}$ is the corresponding parameter set.
The  DAE   is a type of autoencoder that learns  to produce original denoised samples from the inputs contaminated by noise. In an DAE, the parameter set $\theta$ and $ \boldsymbol \theta'$ are trained to minimize the  reconstruction error \cite{zhang2017survey}
\begin{equation}
\small
{\boldsymbol \theta}^{*},{\boldsymbol \theta'}^{*} =\underset{{\boldsymbol \theta} ,{\boldsymbol \theta'} }{\operatorname{argmin} L} \left( \mathbf x,  d_{{\boldsymbol \theta'}}\left(  e_{{\boldsymbol \theta}}  \left(\mathbf x \right) \right)  \right),
\end{equation}
where $L$ is a loss function, such as the squared error loss $L \left(  \mathbf x,    \mathbf {\hat x}\right)  = {\parallel \mathbf x -  \mathbf {\hat x}  \parallel}^2$. Another commonly used loss function is  the cross-entropy loss   $ { {L_{\text{CE}}}} \left(  \mathbf x,    \mathbf {\hat x}\right)= -\sum _{d=1}^{D}  x_{d}\log (\hat x_d )$,
 where $D$ is the length  of the output  vector, $x_d \in \mathbf{x}$ and  $\hat x_d \in \mathbf {\hat x}$.  Note that for cross-entropy loss,  $ \mathbf x$ and $\mathbf {\hat x} $ are in the form of  the bit vector  and bit probability, respectively.
\begin{figure}[]
  \setlength{\abovecaptionskip}{2pt}
  \setlength{\belowcaptionskip}{-12pt}

\centering
\includegraphics[width=0.5\textwidth]{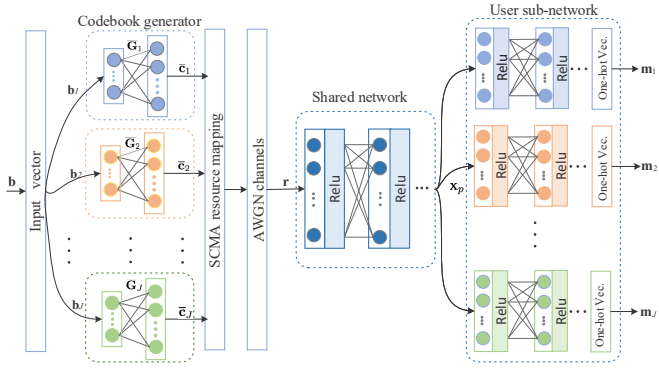}
\caption{The system structure of the proposed E2E-SCMA. }
\label{ae-scma}
\end{figure}
\setlength{\abovedisplayskip}{0pt}

\subsection{Signal Model Inspired Encoder Design }
In our proposed  E2E-SCMA, the mapping from the $j$th  data stream to the $j$th user's constellation, i.e., ${\mathbf c}_{j}=g_{j}(\mathbf {b}_{j})$  is implemented with neural networks.
Note that the SCMA encoding in (\ref{design})  has the same expression with neural network  in (\ref{outl}) when the activation function is linear with  basis $\boldsymbol {z}= \boldsymbol 0^{T}$. Therefore, the codebook generation process, i.e., $g_{j}$, can be implemented with a simple neural network, which only consists of the input layer and output layer.   The weight matrix in the neural network is equivalent to the generator matrix $\mathbf G_j$.  Since the proposed  network   operates  in real domain, the output is   separated into real and imaginary parts. Hence,  (\ref{design})  is re-written as
\begin{equation}
\label{gb}
\small
\mathbf {\bar{c}}_{j} =  {{\mathbf{\bar{G}}}_{j}}{{\mathbf{b}}_{j}},
\end{equation}
with
\begin{equation}
\label{gb1}
   \mathbf {\bar{c}}_{j} = {\left[\begin{array}{c} \Re ({\mathbf {c}_j}) \\ \Im ({\mathbf {c}_j}) \end{array}\right]},
\mathbf {\bar{G}}_j  = {\left[\begin{array}{cc}   ({\mathbf {G}_j^{\text{ R}}})^T  &    ({\mathbf {G}_j^\text{I}} )^T  \end{array}\right]^T},
\end{equation}
  where ${\mathbf {G}_j^\text{R}}$ and $ {\mathbf {G}_j^\text{I}}$ are the generator matrices  of the real and imaginary parts, respectively. Based on the above analysis, the proposed  model based E2E-SCMA with $J$ users is shown in  Fig. \ref{ae-scma}, where the proposed E2E-SCMA is composed of $J$ codebook generators,  a signature mapping module, a channel module, and a multi-user detection module.
The structure of codebook  generator  is inspired by the signal model and  only consists two layers, i.e., the input layer and the output layer.  In addition, the number of nodes for input layer and output layer are $\text{log}_2(M)$ and $2N$, respectively.

In the forward-propagation phase, source message   vector $\mathbf{{b}}_{j}$ first flows  through  codebook generator network, parameterized by $\mathbf {\bar{G}}_j $ to derive the multi-dimensional complex symbol $\mathbf {\bar{c}}_{j}$, and then the symbols are mapped to SCMA resources according to $\mathbf {V}_{j}$. After that, $J$ users' data symbols  are   superimposed before passing  through a Gaussian channel\footnote{In this paper, we focus on  the Gaussian channel case as in \cite{kim2018deep,lin2020novel,Minsig} in order to  give a clear comparison with other benchmarks. The fading channel will be   investigated in future work.}. Finally, the superimposed signal is decoupled to accurately recover source messages based on task-specific sub-networks in the decoder, which will be elaborated in the next subsection.

\subsection{Decoder Design with Multi-task Learning}
At the decoder part, deep multi-task learning is adopted to design the multi-user detector.
The proposed decoder consists of one shared network and  $J$ user specific sub-networks, where the  shared network  is designed for exchanging the information  between the subcarriers and the $j$th task is responsible for  recovering the  $j$th user’s  data.  We employ one-hot vector to represent the input binary message vector  $\mathbf b_j$, namely, each message  $ \mathbf b_{j,m},  m \in \{1,2, \ldots,M\}$ is represented by an  $M$-dimensional one-hot vector ${\boldsymbol {\mathfrak {m}}}_j^m$, which is the $m$th column of the identity matrix $\mathbf I_M$.
 For example, for  $M=4$, the  one-hot mapping  is defined as
\begin{equation}
\small
\begin{aligned}
\mathbf{b}_{ j, 1  }= \left [-1, -1 \right ]^{T}  \leftrightarrow & \boldsymbol {\mathfrak {m}}_j^1=[1,0,0,0],   \\ \mathbf {b}_{ j, 2} = \left [-1,  +1 \right ]^{T} \leftrightarrow & \boldsymbol {\mathfrak {m}}_j^2=[0,1,0,0],   \\ \mathbf {b}_{ j, 3} = \left [+1,  -1 \right ]^{T}  \leftrightarrow & \boldsymbol {\mathfrak {m}}_j^3=[0,0,1,0],   \\ \mathbf {b}_{ j, 4} = \left [ +1,  +1 \right ]^{T}  \leftrightarrow & \boldsymbol {\mathfrak {m}}_j^4=[0,0,0,1].
\end{aligned}
\end{equation}

The decoder can be expressed  as  $d_{{\overline{ { \boldsymbol \theta}}}}^{\text {P}}  d_{{ {\boldsymbol \theta_j}}}^{\text {U}}: {\boldsymbol { r}_j} \to {{\boldsymbol{{ p}}}}_{j}$, where  $d_{{\overline{ { \boldsymbol \theta}}}}^{\text {P}}$ and $ d_{{ {\boldsymbol \theta_j}}}^{\text {U}}$  are the non-linear mapping of the forward DNN for  the shared network and   the  $j$th user'   sub-network, respectively.  $ { {{{\boldsymbol p}}}}_{j} $ is the output   messages,    $\overline{ { \boldsymbol \theta}}$ and ${ \boldsymbol {\theta}}_j$ are the  parameter sets of the  shared network and the $j$th user'   sub-network, respectively. In our implementation, we choose fully-connected DNN  with $L_{\text{P}}$ and $L_{\text{U}}$ layers for both shared network and user sub-network.  The above process can be expressed as
\begin{equation}
\label{decoder}
\small
\begin{aligned}
{ {\boldsymbol{{\mathbf p}}}}_{j}  & = d_{{ {\boldsymbol \theta_j}}}^{\text {U}} \left ({{\mathbf {x}}_{\text {P}}}\right) =\sigma _{j, L_\text {U}}^{\text{U}} \Big ({\mathbf{W}}_{j}^{(L_{\text {U}})}\Big (\sigma _{j,L_\text{U}-1}^{  \text{U} } \cdots \\
&\quad { \sigma _{j,1}^{ \text{U} }\left ({{{ \mathbf W_{j}}^{(1) }}{{\mathbf x}_{\text{P}}}+{{ \mathbf {z}_{j}}^{(1) }}}\right)\cdots + {\mathbf{z}}_{j}^{(L_{\text {U}}-1)}\Big)+{ { \mathbf {z}_{j}}^{\left ({L_{\text {U}}}\right)}}\Big),}  \\
{ {\mathbf{x}}}_{\text{P}}  &=
d_{{\overline{ { \boldsymbol \theta}}}}^{\text {P}} \left ({{ {\mathbf r}}}\right) =\sigma _{L_\text{P}}^{\text{P}} \Big ({\overline{\mathbf{W}}}^{(L_{\text {P}})}\Big ( \sigma _{L_\text{P}-1}^{\text{P}}  \cdots \\
&\quad { \sigma _{1}^{\text{P}} \left ({{ \overline{ {\mathbf W}}^{(1) }}{\mathbf r_j}+{\overline{{ \mathbf z}}^{(1) }}}\right)\cdots +{\overline{\mathbf{z}}}^{(L_{\text {P}}-1)}\Big)+{\overline{{ \mathbf z}}^{\left ({L_{\text {P}}}\right)}}\Big),}
\end{aligned}
\end{equation}
where ${\mathbf{x}}_{\text{P}}$ is the output of   the  shared layer, $\sigma _{l}^{ \text{P}}$  and  $\sigma _{j, l}^{ \text{U}}$ denote the activation function of the $l$th layer of shared network and the $j$th sub-network, respectively. ${{ {\boldsymbol {\theta}}}_{j}}=\Big \lbrace {\mathbf{W}}_{j}^{(1) },{\mathbf{z}}_{j}^{(1) },\ldots,{\mathbf{W}}_{j}^{(L_{\text {U}})},{\mathbf{z}}_{j}^{(L_{\text {U}})}\Big \rbrace, $ and ${\overline{ {\boldsymbol {\theta}}}}=\Big \lbrace {\overline{\mathbf{W}}}^{(1) },$ ${\overline{\mathbf{z}}}^{(1) }, \ldots,{\overline{\mathbf{W}}}^{(L_{\text {\text {P}}})},{\overline{\mathbf{z}}}^{(L_{\text {P}})}\Big \rbrace$ are the parameters to be learned.

Observing that the task of  SCMA detection  is to recover the source messages in a limited search space, such a  problem is equivalent to a typical classification problem in the machine learning field. Hence, this motivates us  to employ the widely used softmax activation for output layer. To facilitate the network convergence, ReLU activation function is adopted for hidden layers.
Assume that  the input of softmax is a vector $\mathbf w_j$ of dimension $M$, and $w_{j,m}$ is the $m$th entry of $\mathbf w_j$. Then, the   softmax activation function takes the following expression:
\begin{equation}
p_{j,m}=\frac{\exp (w_{j,m}) } {\sum _{m' =1}^{M}\exp (w_{j, m' })},
\end{equation}
where $p_{j,m}$ is the  $m$th entry of the output ${ {\mathbf{ p}}}_{j}$ with  $\sum _{m=1}^{M} p_{j,m}=1$.  All hidden layers adopt ReLU activation function, which can facilitate the network convergence during the training process.  As for the loss function, we consider the corresponding softmax cross-entropy  loss for each user. Let ${\mathbf { p}} =  [ {\mathbf {p}}_1^T, { {\mathbf p}}_2^T, \dots, { {\mathbf p}}_J^T  ]^T$ and $\boldsymbol{\mathfrak { m}}   =[ \boldsymbol{\mathfrak { m}}_1^T, \boldsymbol{\mathfrak { m}}_2^T, \dots, \boldsymbol{\mathfrak { m}}_J^T]^T $, where $\boldsymbol{\mathfrak { m}}_j$ is the one hot representation of $\mathbf{b}_j$. The overall loss function is the summation over $J$ users, which can be expressed as
\begin{equation}
\label{loss}
\small
   { {L}}^{\text{E2E-SCMA}}({ {\mathbf p}}, \boldsymbol{\mathfrak {m}})= -\sum _{j=1}^{J}\sum _{m=1}^{M} q_{j,m}\log (p_{j,m}),
 \end{equation}
 where $q_{j,m}$ denotes the $m$th entry of $\boldsymbol {\mathfrak {m}}_j$.   The loss function  measures the difference between predicted probability $\mathbf p$ diverges from the actual label $\boldsymbol {\mathfrak {m}}$. Therefore,  we aim to seek the model parameters $\overline {\mathbf G}_j,  \overline{ { \boldsymbol \theta}}, { \boldsymbol {\theta}}_j$  to minimize  the  overall loss:
\begin{equation}
\small
\{{\overline {\mathbf G}_j}^*,   \overline{ {\boldsymbol \theta}}^*, {{ {\boldsymbol \theta}}_{j}^*}\}=
\underset{  {\left[ {\overline {\mathbf G}_j} \right]_{j=1}^{J}}, \overline{ {\boldsymbol \theta}}, {\left[ {{ {\boldsymbol \theta}}_{j}} \right]_{j=1}^{J}} }  {\mathop{\arg \min }}\,    { {L}}^{\text{E2E-SCMA}}({ {\mathbf p}}, \boldsymbol{\mathfrak {m}}).
 \end{equation}

 \begin{algorithm}[t]
\caption{Training of E2E-SCMA.}
\label{algorithm 1}
\begin{algorithmic}[1]
\REQUIRE{Set $J$, $K$, $\mathbf V_j$, $\alpha_0,$ $ \beta,$ $D$, ${{E}_{b} / {N}_{0}}_{\min},$ ${{E}_{b} / {N}_{0}}_{\max}$, $I_T$ and initialize the network
parameters  ${ {\overline {\mathbf G}_j}},\overline{ {\boldsymbol \theta}},$ ${ {\boldsymbol \theta_j}}, j \in \{ 1, 2, \dots, J\}$.  }\\
\STATE  \textbf{repeat} \\
\STATE  $t\gets 1 $\\
\STATE   Randomly generate training samples $\mathbf b_j$ and transfer $\mathbf b_j$  to one-hot vector ${\boldsymbol {\mathfrak {m}}}_j$\\
\STATE  \textbf{ Froward Propagation} \\
\STATE    $\text{SNR} \gets  \mathcal U \left ({{E}_{b} / {N}_{0}}_{\min}, {{E}_{b} / {N}_{0}}_{\max}\right) $, $\alpha_t \gets \alpha_0  \beta^{\left(   t/D \right) }$
\STATE   $ \mathbf {\bar{c}}_{j} \gets \mathbf b_j$     according to (\ref{gb}) and (\ref{gb1})
\STATE  $ \mathbf {r} \gets$ Obtain $\mathbf {r}$ after resource mapping and pass channel  \\
\STATE $    {{\boldsymbol{{ p}}}}_{j} \gets  d_{{\overline{ { \boldsymbol \theta}}}}^{\text {P}} {\cdot } d_{{ {\boldsymbol \theta_j}}}^{\text {U}} \left( {\boldsymbol { r}} \right)$ according to (\ref{decoder})\\
\STATE    ${ {L}^t}_{batch} \gets  { {L}}^{\text{E2E-SCMA}}({ {\mathbf p}}, \boldsymbol{\mathfrak {m}})$ according to (\ref{loss})
\STATE  \textbf{ Backward  Propagation }\\
\STATE ${\overline {\mathbf G}_j}$, $\boldsymbol{ \theta}_j, \boldsymbol{ {\overline {\theta}}}  \gets  $ Update  $\boldsymbol{ \theta}_j, \boldsymbol{ {\overline {\theta}}}$ with $\alpha_t,$     $\nabla_{{\boldsymbol {\theta}}_j, \mathbf{\overline {\boldsymbol \theta}}} { {L}^t}_{batch}$,  and ${\overline {\mathbf G}_j}$ with $\alpha_t $ and $ \nabla_{{\overline {\mathbf G}_j}, \overline {\boldsymbol \theta}, {\boldsymbol{ \theta}}_j }    { {L}^t}_{batch}$
 with gradient-based optimizer\\
\STATE  $t\gets t+1 $ \\
\STATE  \textbf{until} reaching the maximum iteration number $I_T$
\end{algorithmic}
   \vspace{-0.5em}
\end{algorithm}

\subsection{Training Algorithm}
 The encoder and decoder are jointly optimized with gradient decent based method using forward and backward propagation, such as adaptive moment estimation (ADAM).  \textbf{Algorithm 1} demonstrates the detailed training of the proposed E2E-SCMA system.   $\mathbf {\bar{G}}_j, j=1,2, \ldots, J$ are first initialized  with Huawei codebook  \cite{huawei}. Specifically, we first obtain  $\mathbf {{G}}_j =  \mathbf C_j   \mathbf B^T (\mathbf B\mathbf B^T) ^{-1}$,  where $\mathbf C_j$ denotes  the $j$th user's  codebook in  \cite{huawei} by removing the zero dimensions. Then,  $\mathbf {\bar{G}}_j$ is obtained by concatenating the real and imaginary parts of $\mathbf {{G}}_j$. The weights of the decoder, i.e., $\boldsymbol{ \bar{\theta}}$, and $ \boldsymbol \theta_j$,  are initialized  with  a normal distribution with mean $0$ and variance $1$.  In the forward propagation, the randomly generated input data first flows through the encoder and decoder to obtain an estimation  of the input message. Then, during the backward propagation, the parameters ${ {\overline {\mathbf G}_j}},\overline{ {\boldsymbol \theta}},$ and ${ {\boldsymbol \theta_j}}$ are updated by minimising the total loss.
 In addition, the learning rate $\alpha_t$  decays exponentially at each iteration $t$  with a decay factor of $\beta$ and decay step of $D$.
With respect to the training ${E}_{b} / {N}_{0}$, the authors in \cite{kim2018deep,lin2020novel,Minsig,9310294}    obtained  SCMA codebooks by training the system at a fixed ${E}_{b} / {N}_{0}$. However, in our implementation, the training SNR for each iteration  was randomly generated so that the SNR will be uniformly distributed on $ \mathcal U \left ( {{E}_{b} / {N}_{0}}_{\min}, {{E}_{b} / {N}_{0}}_{\max} \right)$.    This approach allows us to train an SCMA system to work over a wide range of SNR values while maintaining a low error rate performance.

   \subsection{Complexity Analysis}
   The main differences between   E2E-SCMA  and convention SCMA in terms of complexity is the decoder part, i.e., DNN decoder and MPA. Hence, we main focus on analyze the complexity of DNN decoder and MPA. The complexity of MPA is given by $\mathcal{O}\left( {{N}_{iter}}Kd_{f}^{2}{{M}^{df}} \right)$ \cite{yang2016low}, where ${{N}_{itr}}$ is defined as the iteration number   of MPA.  For  E2E-SCMA, we are concerned about  the complexity of online deployment. The main computation in   E2E-SCMA  is matrix multiplication, which is dominated by the two consecutive layers with the largest number of neural nodes. Therefore, we can simply the computation  complexity as $\mathcal{O}\left( L_1L_2 \right)$, where $L_{1}$ and $L_{2}$ are the largest number of neural nodes  of two consecutive layers.

\section {Numerical   results}
 In this section, we evaluate the error rate performance of the proposed E2E-SCMA system in Gaussian channel.  The following  indicating matrix    with $J=6, K=4, N=2$ is   given by
\begin{equation} \label{factor_46}
\begin{aligned}
\mathbf {F_{4 \times 6}}=\left [{
\begin{matrix}
0 &\quad 1 &\quad 1 &\quad 0 &\quad 1 &\quad 0 \\
1 &\quad 0 &\quad 1 &\quad 0 &\quad 0 &\quad 1 \\
0 &\quad 1 &\quad 0 &\quad 1 &\quad 0 &\quad 1 \\
1 &\quad 0 &\quad 0 &\quad 1 &\quad 1 &\quad 0 \\
\end{matrix} }\right].
\end{aligned}
\end{equation}

 The initial learning rate,  decay step and decay factor are set to be  $\alpha_0 = 0.001$, $D = 500$  and $\beta=0.9$, respectively. The batch size for each iteration is set to be  $1000$  for  a trade-off between convergence rate and computational efficiency.  The maximum iteration number is $I_T = 2000$. Therefore, the total number of training samples is $2\times 10^{6}$.
 We choose a wide range of training ${{E}_{b} / {N}_{0}}$, specifically, we set  ${{E}_{b} / {N}_{0}}_{\min}=5$ dB and ${{E}_{b} / {N}_{0}}_{\max}= 11$ dB.
  The  codebook generator is implemented with $\log_2(M)$ input nodes and $2 \times K$ output nodes. For the decoder, the number of  nodes and hidden layers for  shared network are $\{128, 64\}$ and   $L_{P}=2$, respectively, whereas the two parameters for  user  sub-network are  $\{64, 32, 16\}$    and  $L_{U}=3$, respectively. Therefore, the complexity of the E2E-SCMA is   $\mathcal{O}\left( L_1L_2 \right)$, where $L_{1} = 128$ and $L_{2}=64$.

\begin{figure*}
\minipage{0.32\textwidth}
  \includegraphics[width=1.1\textwidth]{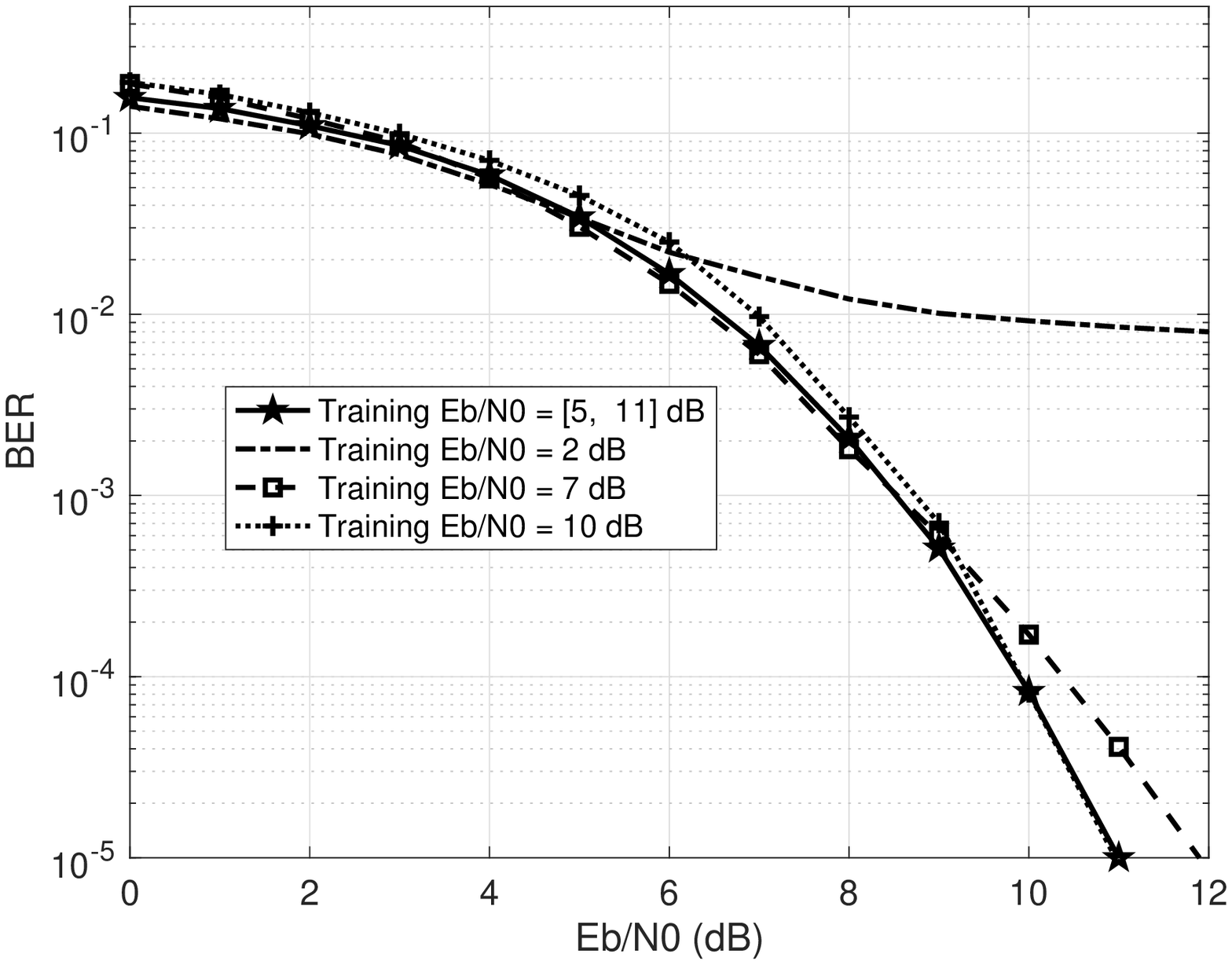}
\caption{System performance of E2E-SCMA trained by various $E_b/N_0  $.  }
\label{Training}
\endminipage\hfill
\minipage{0.32\textwidth}
\includegraphics[width=1.1 \textwidth]{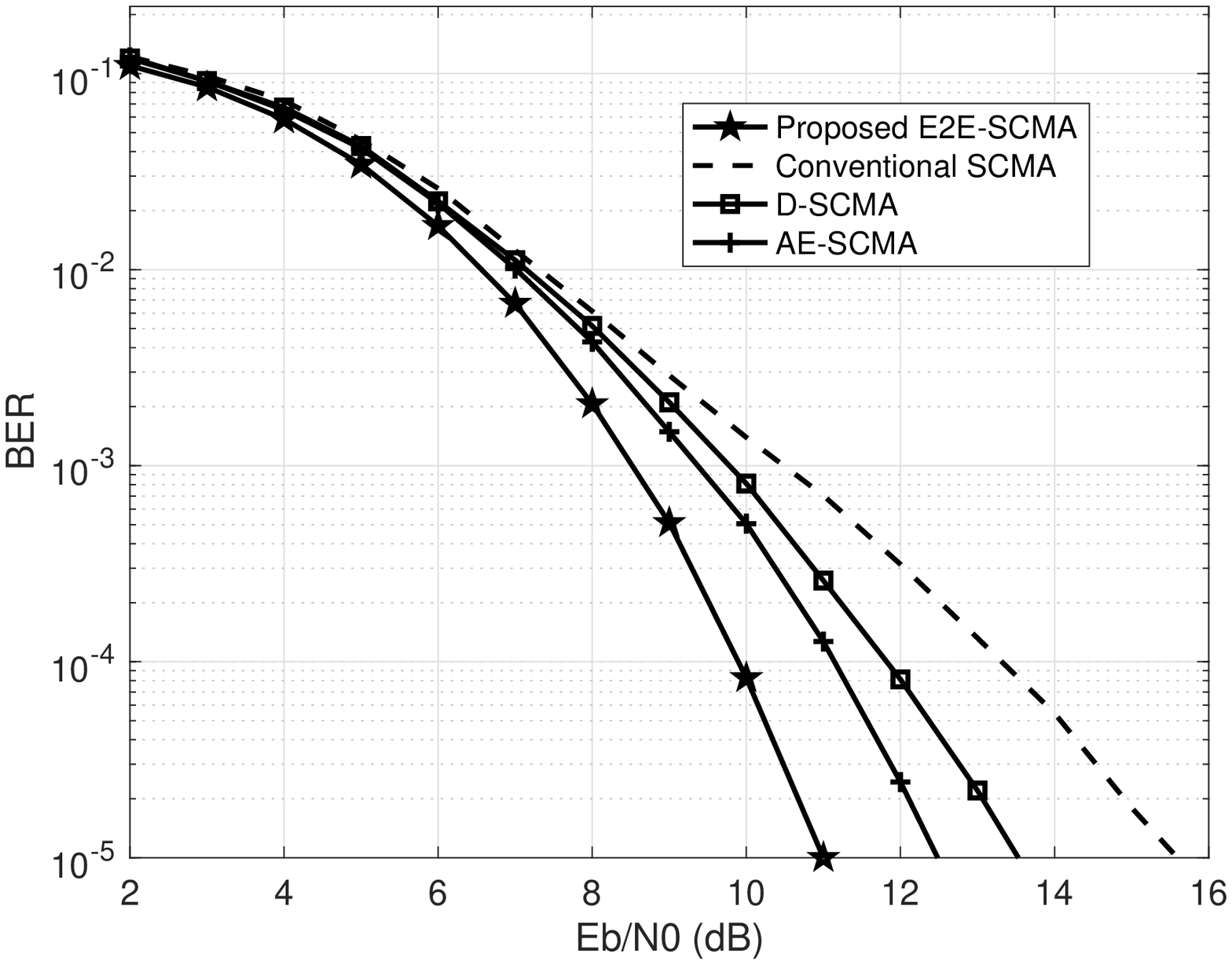}
 \caption{BER performance comparison with different decoders. }
\label{ser}
\endminipage\hfill
\minipage{0.32\textwidth}%
\includegraphics[width=1.1 \textwidth]{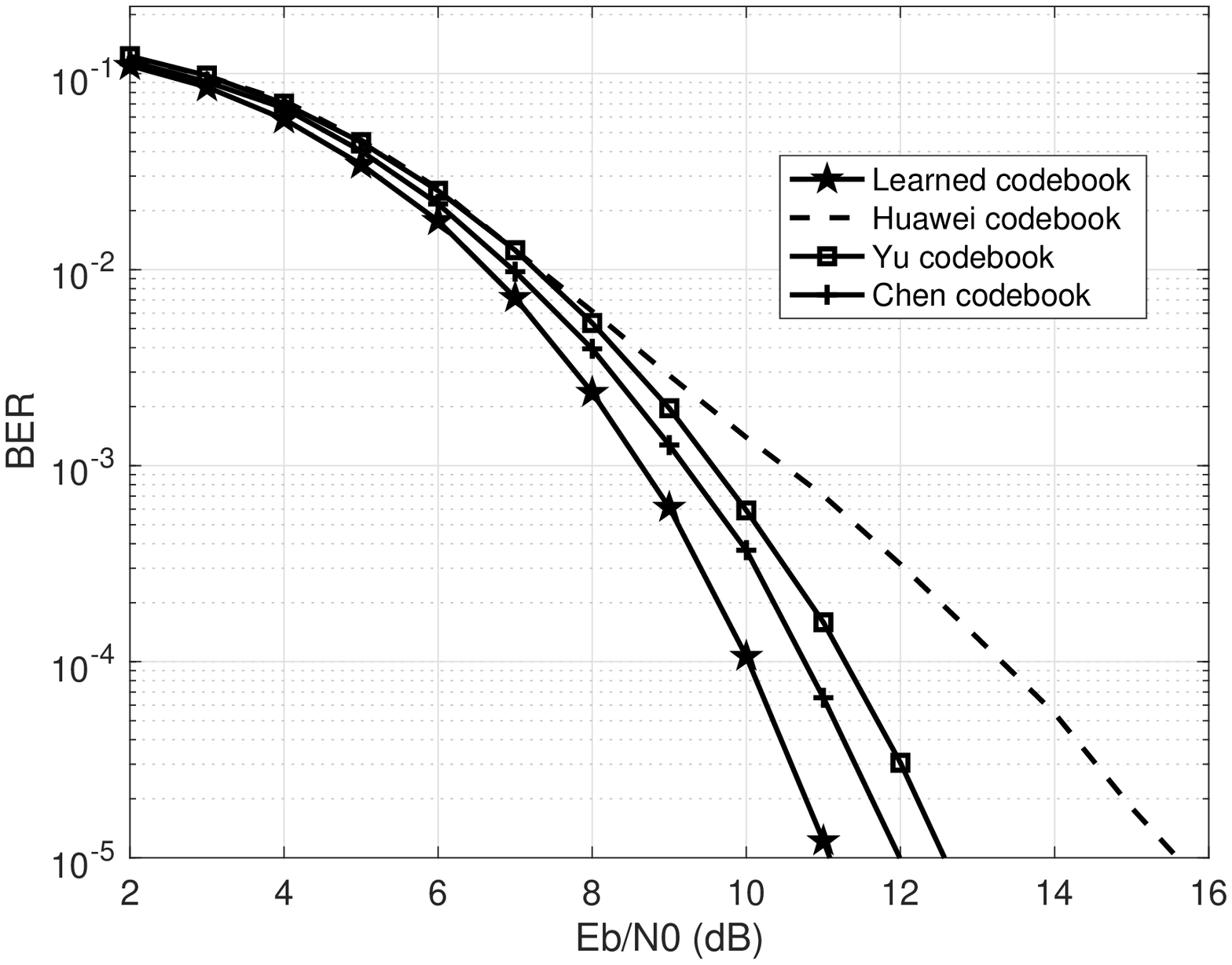}
 \caption{The BER performance of different codebooks with MPA decoder. }
\label{ber}
\endminipage
\end{figure*}

  Since the values of  $E_b/N_0$ in training  influence the BER performance, we  investigate how training samples generated by different $E_b/N_0$  can affect the system performance in Fig.  \ref{Training}. We first train    the system at the fixed $E_b/N_0$ values,  which were set to be   $E_b/N_0= 2 $ dB, $7$ dB and $10$ dB, respectively. Then, the system was also trained in the $E_b/N_0 $ range $ \mathcal U \left( 5, 11  \right)$ dB. It is clearly shown that the low $E_b/N_0$ trained network only performs well in the low $E_b/N_0$  range, whereas the high $E_b/N_0$ trained network  will degrade  the performance  in the low $E_b/N_0$  range. A better way is to  train the network in a wide $E_b/N_0$  range, thus the trained system can harvest the good performance over a wide range $E_b/N_0$ .

 In Fig. \ref{ser}, we compare the BER performance of the proposed   E2E-SCMA  scheme with the AE-SCMA scheme \cite{lin2020novel}, the D-SCMA scheme  \cite{kim2018deep},  and the   conventional SCMA scheme with Huawei codebook \cite{huawei}. The MPA decoder is employed for conventional SCMA scheme  to compare with deep learning  designed SCMA system.
The results show  that the proposed scheme significantly outperforms all conventional SCMA schemes.  Specifically, the proposed E2E-SCMA achieves $3.5$ dB gain and $1.8$ dB gain over D-SCMA, AE-SCMA scheme at SER $ = 10^{-5}$, respectively.

 To  evaluate the codebook obtained by E2E-SCMA scheme, we compare the MED and corresponding BER performance with MPA decoder with the state of art codebooks.  The MED is obtain by calculating  ${{{M}^{J}}\left( {{M}^{J}}-1 \right)}/{2}$ mutual distances between ${{M}^{J}}$ superimposed codewords, which constitute a superimposed constellation ${{\Phi } }$.  Hence, the MED can be expressed as
 \begin{equation*}
\min \left\{ {\parallel  \mathbf  v_n -  \mathbf {v}_m  \parallel}^2, \! \forall {{\mathbf{v}}_{n}},{{\mathbf{v}}_{m}}\in \Phi, \! \forall m,n\in {{Z}_{{{M}^{J}}}},m\ne n \right\},
  \end{equation*}
where ${{Z}_{{{M}^{J}}}}$ stands for the integer set $\left\{ 1,2,\ldots ,{{M}^{J}} \right\}$.  Specifically, the MED of learned codebook is compared with  Huawei codebook \cite{huawei}, Chen codebook \cite{chen2020design}   and Yu codebook\cite{yu2015optimized}. The results are presented in Table I.  It can be seen that the learned codebook  owns  MED = $1.17$ and is  higher  than other codebooks. Then, BER comparisons of different codebooks with MPA decoder  are shown in Fig. \ref{ber}.  The proposed codebook achieves $4.8$ dB gain over the Huawei codebook at BER $ = 10^{-5}$,  about $1.8$ dB gain over the Yu codebook, and $1$ dB gain over the Chen codebook at BER $ = 10^{-5}$.  The proposed codebook and the codebooks employed for comparison   are all available at our GuitHub project\footnote{\url{https://github.com/ethanlq/SCMA-codebook/tree/main/CB_autoencoder} }.

 \vspace{-0.5em}
  \section{Conclusion}
 In this paper,  we  have proposed  an E2E-SCMA by joint  optimization of SCMA encoder and decoder with the aid of DAE.  Our key idea is to design the SCMA encoder by taking into account of the mapping procedure and then optimize the decoder with multi-task learning approach.  Simulation results showed that the use of   multi-task learning  technique enables efficient derivation of codebook and decoding strategy for a sparse and multidimensional superimposed signal. In   addition, our  proposed scheme outperforms conventional schemes and existing autoencoder SCMA in terms of both error rate and computational complexity.
\begin{table}
\begin{center}
\caption{\label{table} A comparison of MEDs of different codebooks }
\begin{tabular}{lcl}
\toprule  Codebook    &      MED  \\
\midrule   Huawei  \cite{huawei}    &      0.56  \\
           Yu   \cite{yu2015optimized}   &     0.90  \\
           Chen  \cite{chen2020design}   &     1.07  \\
Learned codebook     &    1.17 \\
\bottomrule
 \end{tabular}
\end{center}
\vspace{-2em}
 \end{table}



\ifCLASSOPTIONcaptionsoff
  \newpage
\fi

\bibliographystyle{IEEEtran}
  \bibliography{ref}

\end{document}